\PassOptionsToPackage{hyphens}{url}
\documentclass[format=sigconf,letterpaper,10pt,authorversion]{acmart}
\usepackage{xurl}
\usepackage[hyphens]{url}
\usepackage{hyperref}
\usepackage{booktabs} %

\usepackage{tabularx}
\usepackage{listliketab}

\usepackage{etoolbox}
\usepackage[american]{babel}

\hypersetup{breaklinks=true}
\usepackage{epsfig}
\usepackage{caption}
\usepackage{subcaption}
\usepackage{ulem}

\usepackage{colortbl}
\usepackage{units}

\usepackage[inline]{enumitem}
\usepackage{balance}
\usepackage{nth}
\usepackage{dblfloatfix}
\usepackage{listings}
\usepackage{tikz}
\usetikzlibrary{automata,backgrounds,fit,positioning,shapes,calc,arrows.meta}

\hypersetup{pdfstartview=FitH,pdfpagelayout=SinglePage}

\newcommand{\eg}{e.g.,}
\newcommand{\etal}{et~al\@ifnextchar.{}{.\@}}
\newcommand{\etc}{etc\@ifnextchar.{}{.\@}}

\newcommand{\sref}[1]{Section~\ref{#1}}
\newcommand{\fig}[1]{Figure~\ref{#1}}
\newcommand{\tab}[1]{Table~\ref{#1}}

\newcommand{\afblock}[1]{\noindent{\textbf{#1}}}

\makeatletter
\DeclareRobustCommand{\textsupsub}[2]{{%
  \m@th\ensuremath{%
    ^{\mbox{\fontsize\sf@size\z@#1}}%
    _{\mbox{\fontsize\sf@size\z@#2}}%
  }%
}}
\makeatother

\copyrightyear{2019}
\acmYear{2019}
\setcopyright{acmlicensed}
\acmConference[ANRW '19]{ANRW '19: Applied Networking Research
Workshop}{July 22, 2019}{Montreal, QC, Canada}
\acmBooktitle{ANRW '19: Applied Networking Research Workshop (ANRW '19),
July 22, 2019, Montreal, QC, Canada}
\acmPrice{15.00}
\acmDOI{10.1145/3340301.3341123}
\acmISBN{978-1-4503-6848-3/19/07}

\AtEndDocument{\balance}

\begin{document}
\normalem

\title[A Performance Perspective on Web Optimized\\Protocol Stacks: TCP+TLS+HTTP/2 vs.\ QUIC]{A Performance Perspective on Web Optimized Protocol Stacks: TCP+TLS+HTTP/2 vs.\ QUIC}
\renewcommand{\shortauthors}{}

\author{Konrad Wolsing} %
\author{Jan R\"uth}
\author{Klaus Wehrle}
\author{Oliver Hohlfeld}
\authornote{Is now at Brandenburg University of Technology}
\affiliation{
\institution{RWTH Aachen University, Germany}
}
\email{{wolsing, rueth, wehrle, hohlfeld}@comsys.rwth-aachen.de}
\renewcommand{\shortauthors}{Wolsing and R\"uth, et al.}

\begin{abstract}
Existing performance comparisons of QUIC and TCP compared an optimized QUIC to an unoptimized TCP stack.
By neglecting available TCP improvements inherently included in QUIC, comparisons do not shed light on the performance of current web stacks.
In this paper, we can show that tuning TCP parameters is not negligible and directly yields significant improvements.
Nevertheless, QUIC still outperforms even our tuned variant of TCP.
This performance advantage is mostly caused by QUIC's reduced RTT design during connection establishment, and, in case of lossy networks due to its ability to circumvent head-of-line blocking.
\end{abstract}


\maketitle
\def\MeanDiffFvcDaagcQuicQuicbbr{209.8} %
\def\MeanDiffFvcDaagcTcpbbrQuic{-1101.0} %
\def\MeanDiffFvcDaagcTcpbbrQuicbbr{-891.2} %
\def\MeanDiffFvcDaagcTcppQuic{-1183.6} %
\def\MeanDiffFvcDaagcTcppTcpbbr{-82.6} %
\def\MeanDiffFvcDaagcTcpQuic{-1298.1} %
\def\MeanDiffFvcDaagcTcpQuicbbr{-1088.4} %
\def\MeanDiffFvcDaagcTcpTcpbbr{-197.1} %
\def\MeanDiffFvcDaagcTcpTcpp{-114.5} %
\def\MeanDiffFvcDslQuicQuicbbr{5.0} %
\def\MeanDiffFvcDslTcpbbrQuic{-64.1} %
\def\MeanDiffFvcDslTcpbbrQuicbbr{-59.1} %
\def\MeanDiffFvcDslTcppQuic{-45.6} %
\def\MeanDiffFvcDslTcppTcpbbr{18.4} %
\def\MeanDiffFvcDslTcpQuic{-84.5} %
\def\MeanDiffFvcDslTcpQuicbbr{-79.6} %
\def\MeanDiffFvcDslTcpTcpbbr{-20.5} %
\def\MeanDiffFvcDslTcpTcpp{-38.9} %
\def\MeanDiffFvcLteQuicQuicbbr{3.2} %
\def\MeanDiffFvcLteTcpbbrQuic{-164.3} %
\def\MeanDiffFvcLteTcpbbrQuicbbr{-161.1} %
\def\MeanDiffFvcLteTcppQuic{-174.0} %
\def\MeanDiffFvcLteTcppTcpbbr{-9.7} %
\def\MeanDiffFvcLteTcpQuic{-289.8} %
\def\MeanDiffFvcLteTcpQuicbbr{-286.6} %
\def\MeanDiffFvcLteTcpTcpbbr{-125.5} %
\def\MeanDiffFvcLteTcpTcpp{-115.8} %
\def\MeanDiffFvcMssQuicQuicbbr{-15.6} %
\def\MeanDiffFvcMssTcpbbrQuic{-1711.9} %
\def\MeanDiffFvcMssTcpbbrQuicbbr{-1727.5} %
\def\MeanDiffFvcMssTcppQuic{-4578.1} %
\def\MeanDiffFvcMssTcppTcpbbr{-2866.2} %
\def\MeanDiffFvcMssTcpQuic{-6699.9} %
\def\MeanDiffFvcMssTcpQuicbbr{-6715.5} %
\def\MeanDiffFvcMssTcpTcpbbr{-4988.0} %
\def\MeanDiffFvcMssTcpTcpp{-2121.8} %
\def\MeanDiffPltDaagcQuicQuicbbr{1474.1} %
\def\MeanDiffPltDaagcTcpbbrQuic{-4417.7} %
\def\MeanDiffPltDaagcTcpbbrQuicbbr{-2943.6} %
\def\MeanDiffPltDaagcTcppQuic{-7405.6} %
\def\MeanDiffPltDaagcTcppTcpbbr{-2987.9} %
\def\MeanDiffPltDaagcTcpQuic{-6935.1} %
\def\MeanDiffPltDaagcTcpQuicbbr{-5461.0} %
\def\MeanDiffPltDaagcTcpTcpbbr{-2517.4} %
\def\MeanDiffPltDaagcTcpTcpp{470.5} %
\def\MeanDiffPltDslQuicQuicbbr{-25.6} %
\def\MeanDiffPltDslTcpbbrQuic{-62.5} %
\def\MeanDiffPltDslTcpbbrQuicbbr{-88.1} %
\def\MeanDiffPltDslTcppQuic{-102.2} %
\def\MeanDiffPltDslTcppTcpbbr{-39.7} %
\def\MeanDiffPltDslTcpQuic{-182.3} %
\def\MeanDiffPltDslTcpQuicbbr{-208.0} %
\def\MeanDiffPltDslTcpTcpbbr{-119.8} %
\def\MeanDiffPltDslTcpTcpp{-80.1} %
\def\MeanDiffPltLteQuicQuicbbr{-35.0} %
\def\MeanDiffPltLteTcpbbrQuic{-228.1} %
\def\MeanDiffPltLteTcpbbrQuicbbr{-263.1} %
\def\MeanDiffPltLteTcppQuic{-261.6} %
\def\MeanDiffPltLteTcppTcpbbr{-33.5} %
\def\MeanDiffPltLteTcpQuic{-426.1} %
\def\MeanDiffPltLteTcpQuicbbr{-461.1} %
\def\MeanDiffPltLteTcpTcpbbr{-198.0} %
\def\MeanDiffPltLteTcpTcpp{-164.5} %
\def\MeanDiffPltMssQuicQuicbbr{-14701.7} %
\def\MeanDiffPltMssTcpbbrQuic{11395.4} %
\def\MeanDiffPltMssTcpbbrQuicbbr{-3306.3} %
\def\MeanDiffPltMssTcppQuic{-26086.5} %
\def\MeanDiffPltMssTcppTcpbbr{-37481.9} %
\def\MeanDiffPltMssTcpQuic{-29832.9} %
\def\MeanDiffPltMssTcpQuicbbr{-44534.6} %
\def\MeanDiffPltMssTcpTcpbbr{-41228.3} %
\def\MeanDiffPltMssTcpTcpp{-3746.4} %
\def\MeanDiffSiDaagcQuicQuicbbr{386.7} %
\def\MeanDiffSiDaagcTcpbbrQuic{-1759.2} %
\def\MeanDiffSiDaagcTcpbbrQuicbbr{-1372.5} %
\def\MeanDiffSiDaagcTcppQuic{-2632.5} %
\def\MeanDiffSiDaagcTcppTcpbbr{-873.3} %
\def\MeanDiffSiDaagcTcpQuic{-2413.0} %
\def\MeanDiffSiDaagcTcpQuicbbr{-2026.3} %
\def\MeanDiffSiDaagcTcpTcpbbr{-653.8} %
\def\MeanDiffSiDaagcTcpTcpp{219.5} %
\def\MeanDiffSiDslQuicQuicbbr{4.2} %
\def\MeanDiffSiDslTcpbbrQuic{-80.5} %
\def\MeanDiffSiDslTcpbbrQuicbbr{-76.3} %
\def\MeanDiffSiDslTcppQuic{-87.1} %
\def\MeanDiffSiDslTcppTcpbbr{-6.6} %
\def\MeanDiffSiDslTcpQuic{-131.3} %
\def\MeanDiffSiDslTcpQuicbbr{-127.2} %
\def\MeanDiffSiDslTcpTcpbbr{-50.8} %
\def\MeanDiffSiDslTcpTcpp{-44.2} %
\def\MeanDiffSiLteQuicQuicbbr{4.0} %
\def\MeanDiffSiLteTcpbbrQuic{-203.7} %
\def\MeanDiffSiLteTcpbbrQuicbbr{-199.7} %
\def\MeanDiffSiLteTcppQuic{-215.9} %
\def\MeanDiffSiLteTcppTcpbbr{-12.2} %
\def\MeanDiffSiLteTcpQuic{-344.9} %
\def\MeanDiffSiLteTcpQuicbbr{-340.9} %
\def\MeanDiffSiLteTcpTcpbbr{-141.2} %
\def\MeanDiffSiLteTcpTcpp{-128.9} %
\def\MeanDiffSiMssQuicQuicbbr{-1828.3} %
\def\MeanDiffSiMssTcpbbrQuic{-263.2} %
\def\MeanDiffSiMssTcpbbrQuicbbr{-2091.5} %
\def\MeanDiffSiMssTcppQuic{-8364.8} %
\def\MeanDiffSiMssTcppTcpbbr{-8101.6} %
\def\MeanDiffSiMssTcpQuic{-10699.1} %
\def\MeanDiffSiMssTcpQuicbbr{-12527.4} %
\def\MeanDiffSiMssTcpTcpbbr{-10435.9} %
\def\MeanDiffSiMssTcpTcpp{-2334.3} %
\def\MeanDiffVcDaagcQuicQuicbbr{590.9} %
\def\MeanDiffVcDaagcTcpbbrQuic{-839.2} %
\def\MeanDiffVcDaagcTcpbbrQuicbbr{-248.3} %
\def\MeanDiffVcDaagcTcppQuic{-2767.1} %
\def\MeanDiffVcDaagcTcppTcpbbr{-1927.9} %
\def\MeanDiffVcDaagcTcpQuic{-2330.2} %
\def\MeanDiffVcDaagcTcpQuicbbr{-1739.2} %
\def\MeanDiffVcDaagcTcpTcpbbr{-1490.9} %
\def\MeanDiffVcDaagcTcpTcpp{436.9} %
\def\MeanDiffVcDslQuicQuicbbr{-1.2} %
\def\MeanDiffVcDslTcpbbrQuic{-91.5} %
\def\MeanDiffVcDslTcpbbrQuicbbr{-92.7} %
\def\MeanDiffVcDslTcppQuic{-110.4} %
\def\MeanDiffVcDslTcppTcpbbr{-18.9} %
\def\MeanDiffVcDslTcpQuic{-150.9} %
\def\MeanDiffVcDslTcpQuicbbr{-152.0} %
\def\MeanDiffVcDslTcpTcpbbr{-59.3} %
\def\MeanDiffVcDslTcpTcpp{-40.4} %
\def\MeanDiffVcLteQuicQuicbbr{18.4} %
\def\MeanDiffVcLteTcpbbrQuic{-202.6} %
\def\MeanDiffVcLteTcpbbrQuicbbr{-184.2} %
\def\MeanDiffVcLteTcppQuic{-272.7} %
\def\MeanDiffVcLteTcppTcpbbr{-70.1} %
\def\MeanDiffVcLteTcpQuic{-391.4} %
\def\MeanDiffVcLteTcpQuicbbr{-373.0} %
\def\MeanDiffVcLteTcpTcpbbr{-188.8} %
\def\MeanDiffVcLteTcpTcpp{-118.7} %
\def\MeanDiffVcMssQuicQuicbbr{-5517.0} %
\def\MeanDiffVcMssTcpbbrQuic{3173.8} %
\def\MeanDiffVcMssTcpbbrQuicbbr{-2343.2} %
\def\MeanDiffVcMssTcppQuic{-11683.3} %
\def\MeanDiffVcMssTcppTcpbbr{-14857.2} %
\def\MeanDiffVcMssTcpQuic{-14388.8} %
\def\MeanDiffVcMssTcpQuicbbr{-19905.8} %
\def\MeanDiffVcMssTcpTcpbbr{-17562.6} %
\def\MeanDiffVcMssTcpTcpp{-2705.4} %
\def\MeanGainFvcDaagcQuicQuicbbr{0.02} %
\def\MeanGainFvcDaagcTcpbbrQuic{-0.16} %
\def\MeanGainFvcDaagcTcpbbrQuicbbr{-0.14} %
\def\MeanGainFvcDaagcTcppQuic{-0.14} %
\def\MeanGainFvcDaagcTcppTcpbbr{0.03} %
\def\MeanGainFvcDaagcTcpQuic{-0.15} %
\def\MeanGainFvcDaagcTcpQuicbbr{-0.14} %
\def\MeanGainFvcDaagcTcpTcpbbr{0.03} %
\def\MeanGainFvcDaagcTcpTcpp{-0.00} %
\def\MeanGainFvcDslQuicQuicbbr{0.01} %
\def\MeanGainFvcDslTcpbbrQuic{-0.10} %
\def\MeanGainFvcDslTcpbbrQuicbbr{-0.10} %
\def\MeanGainFvcDslTcppQuic{-0.08} %
\def\MeanGainFvcDslTcppTcpbbr{0.02} %
\def\MeanGainFvcDslTcpQuic{-0.13} %
\def\MeanGainFvcDslTcpQuicbbr{-0.13} %
\def\MeanGainFvcDslTcpTcpbbr{-0.03} %
\def\MeanGainFvcDslTcpTcpp{-0.06} %
\def\MeanGainFvcLteQuicQuicbbr{0.01} %
\def\MeanGainFvcLteTcpbbrQuic{-0.16} %
\def\MeanGainFvcLteTcpbbrQuicbbr{-0.15} %
\def\MeanGainFvcLteTcppQuic{-0.16} %
\def\MeanGainFvcLteTcppTcpbbr{-0.01} %
\def\MeanGainFvcLteTcpQuic{-0.24} %
\def\MeanGainFvcLteTcpQuicbbr{-0.23} %
\def\MeanGainFvcLteTcpTcpbbr{-0.09} %
\def\MeanGainFvcLteTcpTcpp{-0.09} %
\def\MeanGainFvcMssQuicQuicbbr{0.02} %
\def\MeanGainFvcMssTcpbbrQuic{-0.20} %
\def\MeanGainFvcMssTcpbbrQuicbbr{-0.20} %
\def\MeanGainFvcMssTcppQuic{-0.33} %
\def\MeanGainFvcMssTcppTcpbbr{-0.14} %
\def\MeanGainFvcMssTcpQuic{-0.42} %
\def\MeanGainFvcMssTcpQuicbbr{-0.41} %
\def\MeanGainFvcMssTcpTcpbbr{-0.26} %
\def\MeanGainFvcMssTcpTcpp{-0.13} %
\def\MeanGainPltDaagcQuicQuicbbr{0.03} %
\def\MeanGainPltDaagcTcpbbrQuic{-0.13} %
\def\MeanGainPltDaagcTcpbbrQuicbbr{-0.10} %
\def\MeanGainPltDaagcTcppQuic{-0.16} %
\def\MeanGainPltDaagcTcppTcpbbr{-0.03} %
\def\MeanGainPltDaagcTcpQuic{-0.16} %
\def\MeanGainPltDaagcTcpQuicbbr{-0.13} %
\def\MeanGainPltDaagcTcpTcpbbr{-0.03} %
\def\MeanGainPltDaagcTcpTcpp{-0.00} %
\def\MeanGainPltDslQuicQuicbbr{-0.01} %
\def\MeanGainPltDslTcpbbrQuic{-0.06} %
\def\MeanGainPltDslTcpbbrQuicbbr{-0.08} %
\def\MeanGainPltDslTcppQuic{-0.08} %
\def\MeanGainPltDslTcppTcpbbr{-0.01} %
\def\MeanGainPltDslTcpQuic{-0.14} %
\def\MeanGainPltDslTcpQuicbbr{-0.15} %
\def\MeanGainPltDslTcpTcpbbr{-0.08} %
\def\MeanGainPltDslTcpTcpp{-0.06} %
\def\MeanGainPltLteQuicQuicbbr{-0.00} %
\def\MeanGainPltLteTcpbbrQuic{-0.11} %
\def\MeanGainPltLteTcpbbrQuicbbr{-0.12} %
\def\MeanGainPltLteTcppQuic{-0.12} %
\def\MeanGainPltLteTcppTcpbbr{-0.01} %
\def\MeanGainPltLteTcpQuic{-0.18} %
\def\MeanGainPltLteTcpQuicbbr{-0.19} %
\def\MeanGainPltLteTcpTcpbbr{-0.08} %
\def\MeanGainPltLteTcpTcpp{-0.07} %
\def\MeanGainPltMssQuicQuicbbr{-0.17} %
\def\MeanGainPltMssTcpbbrQuic{0.21} %
\def\MeanGainPltMssTcpbbrQuicbbr{-0.14} %
\def\MeanGainPltMssTcppQuic{-0.38} %
\def\MeanGainPltMssTcppTcpbbr{-0.38} %
\def\MeanGainPltMssTcpQuic{-0.46} %
\def\MeanGainPltMssTcpQuicbbr{-0.55} %
\def\MeanGainPltMssTcpTcpbbr{-0.47} %
\def\MeanGainPltMssTcpTcpp{-0.12} %
\def\MeanGainSiDaagcQuicQuicbbr{0.02} %
\def\MeanGainSiDaagcTcpbbrQuic{-0.12} %
\def\MeanGainSiDaagcTcpbbrQuicbbr{-0.10} %
\def\MeanGainSiDaagcTcppQuic{-0.14} %
\def\MeanGainSiDaagcTcppTcpbbr{-0.00} %
\def\MeanGainSiDaagcTcpQuic{-0.15} %
\def\MeanGainSiDaagcTcpQuicbbr{-0.13} %
\def\MeanGainSiDaagcTcpTcpbbr{-0.00} %
\def\MeanGainSiDaagcTcpTcpp{0.01} %
\def\MeanGainSiDslQuicQuicbbr{0.01} %
\def\MeanGainSiDslTcpbbrQuic{-0.09} %
\def\MeanGainSiDslTcpbbrQuicbbr{-0.09} %
\def\MeanGainSiDslTcppQuic{-0.09} %
\def\MeanGainSiDslTcppTcpbbr{-0.00} %
\def\MeanGainSiDslTcpQuic{-0.13} %
\def\MeanGainSiDslTcpQuicbbr{-0.13} %
\def\MeanGainSiDslTcpTcpbbr{-0.05} %
\def\MeanGainSiDslTcpTcpp{-0.05} %
\def\MeanGainSiLteQuicQuicbbr{0.00} %
\def\MeanGainSiLteTcpbbrQuic{-0.13} %
\def\MeanGainSiLteTcpbbrQuicbbr{-0.13} %
\def\MeanGainSiLteTcppQuic{-0.14} %
\def\MeanGainSiLteTcppTcpbbr{-0.01} %
\def\MeanGainSiLteTcpQuic{-0.21} %
\def\MeanGainSiLteTcpQuicbbr{-0.21} %
\def\MeanGainSiLteTcpTcpbbr{-0.09} %
\def\MeanGainSiLteTcpTcpp{-0.08} %
\def\MeanGainSiMssQuicQuicbbr{-0.05} %
\def\MeanGainSiMssTcpbbrQuic{-0.09} %
\def\MeanGainSiMssTcpbbrQuicbbr{-0.18} %
\def\MeanGainSiMssTcppQuic{-0.36} %
\def\MeanGainSiMssTcppTcpbbr{-0.26} %
\def\MeanGainSiMssTcpQuic{-0.44} %
\def\MeanGainSiMssTcpQuicbbr{-0.48} %
\def\MeanGainSiMssTcpTcpbbr{-0.35} %
\def\MeanGainSiMssTcpTcpp{-0.12} %
\def\MeanGainVcDaagcQuicQuicbbr{0.03} %
\def\MeanGainVcDaagcTcpbbrQuic{-0.03} %
\def\MeanGainVcDaagcTcpbbrQuicbbr{-0.01} %
\def\MeanGainVcDaagcTcppQuic{-0.13} %
\def\MeanGainVcDaagcTcppTcpbbr{-0.02} %
\def\MeanGainVcDaagcTcpQuic{-0.15} %
\def\MeanGainVcDaagcTcpQuicbbr{-0.12} %
\def\MeanGainVcDaagcTcpTcpbbr{-0.02} %
\def\MeanGainVcDaagcTcpTcpp{0.01} %
\def\MeanGainVcDslQuicQuicbbr{0.00} %
\def\MeanGainVcDslTcpbbrQuic{-0.09} %
\def\MeanGainVcDslTcpbbrQuicbbr{-0.09} %
\def\MeanGainVcDslTcppQuic{-0.09} %
\def\MeanGainVcDslTcppTcpbbr{-0.00} %
\def\MeanGainVcDslTcpQuic{-0.13} %
\def\MeanGainVcDslTcpQuicbbr{-0.13} %
\def\MeanGainVcDslTcpTcpbbr{-0.04} %
\def\MeanGainVcDslTcpTcpp{-0.04} %
\def\MeanGainVcLteQuicQuicbbr{0.01} %
\def\MeanGainVcLteTcpbbrQuic{-0.11} %
\def\MeanGainVcLteTcpbbrQuicbbr{-0.11} %
\def\MeanGainVcLteTcppQuic{-0.14} %
\def\MeanGainVcLteTcppTcpbbr{-0.01} %
\def\MeanGainVcLteTcpQuic{-0.20} %
\def\MeanGainVcLteTcpQuicbbr{-0.19} %
\def\MeanGainVcLteTcpTcpbbr{-0.08} %
\def\MeanGainVcLteTcpTcpp{-0.07} %
\def\MeanGainVcMssQuicQuicbbr{-0.07} %
\def\MeanGainVcMssTcpbbrQuic{0.02} %
\def\MeanGainVcMssTcpbbrQuicbbr{-0.17} %
\def\MeanGainVcMssTcppQuic{-0.37} %
\def\MeanGainVcMssTcppTcpbbr{-0.30} %
\def\MeanGainVcMssTcpQuic{-0.45} %
\def\MeanGainVcMssTcpQuicbbr{-0.50} %
\def\MeanGainVcMssTcpTcpbbr{-0.39} %
\def\MeanGainVcMssTcpTcpp{-0.11} %

\section{Introduction}
\label{sec:intro}

The advancement of Web application and services resulted in an ongoing evolution of the Web protocol stack.
Driving reasons are security and privacy or the realization of latency-sensitive Web services.
Today, the typical Web stack involves using HTTP/2 over TLS over TCP, making it practically one (ossified) protocol.
While parts of the protocols have been designed to account for the others, this protocol stacking still suffers from inefficiencies, \eg{} head-of-line blocking.
Even though protocol extensions promise higher efficiency (\eg{} TLS 1.3 early-data~\cite{RFC8446} or TCP Fast Open~\cite{radhakrishnan11:conext:TFO}), the ossification around the initial designs challenges their deployment.

QUIC~\cite{langley17:sigcomm:quic} (as used in HTTP/3) combines the concepts of TCP, TLS, HTTP/2, tightly coupled into a new protocol that enables to utilize cross-layer information and to evolve without ossification.
While it fixes some of TCP's shortcomings like head-of-line blocking when used with HTTP, its design, in the first place, should enable evolution.

A number of studies showed that QUIC outperforms the \textit{classical} TCP-based stack~\cite{biswal16:globecom:does,carlucci15:sac:http,cook17:icc:quic,kakhki17:imc:taking,megyesi16:icc:quick,yu17:ipccc:quic}---that is by comparing QUIC to an unoptimized TCP-based stack; a limitation that we address in this paper.
Current QUIC implementations were specifically designed and parameterized for the Web.
In contrast, stock TCP implementations, as in the Linux kernel, are not specialized and are built to perform well on a large set of devices, networks, and workloads.
However, we have shown~\cite{rueth18:tma:iwcdn} that large content providers fine-tune their TCP stacks (\eg{} by tuning the initial window size) to improve content delivery.
All studies known to us neglect this fact and indeed compare an out-of-the-box TCP with a highly-tuned QUIC Web stack and show that the optimized version is superior.
Furthermore, they often utilize simple Web performance metrics like page load time (PLT) to reason about the page loading speed, even though it is long known that PLT does not correlate to user-perceived speeds~\cite{kelton17:nsdi:webgaze,zimmermann17:inetqoe:qoepush,bocchi17:pam:web}.

In this paper, we seek to close this gap by parameterizing TCP similar to QUIC to enable a fair comparison.
This includes increasing the initial congestion window, enabling pacing, setting no slow start after idle, and tuning the kernel buffers to match QUIC's defaults.
We further enable BBR instead of the CUBIC as the congestion control algorithm in one scenario.
We show that this previously neglected tuning of TCP impacts its performance.
We find that for broadband access, QUIC's RTT-optimized connection establishment indeed increases the loading speed, but otherwise compares to TCP.
If optimizations such as TLS 1.3 early-data or TCP Fast Open were deployed, QUIC and TCP would compare well.
In lossy networks, QUIC clearly outperforms the current Web stack, which we mainly attribute to its ability to progress streams independently of head-of-line blocking.
Our comparison is based on visual Web performance metrics that better correlate to human perception than traditionally used loading times.
To evaluate real-world websites, we extend the Mahimahi framework to utilize the Google QUIC Web stack to perform reproducible comparisons between TCP and QUIC on a large scale of settings.
This work does not raise any ethical issues and makes the following contributions:

\vspace{-0.55em}
\begin{itemize}[noitemsep,topsep=5pt,leftmargin=9pt]
	\item We provide the first study that performs an eye-level comparison of TCP+TLS+HTTP/2 and QUIC.
	\item Our study highlights that QUIC can indeed outperform TCP in a variety of settings but so does a tuned TCP.
	\item Tuning TCP closes the gap to QUIC and shows that TCP is still very competitive to QUIC. 
	\item Our study further highlights the immense impact of choice of congestion control, especially in lossy environments.
	\item We add QUIC support to Mahimahi to enable reproducible QUIC research. It replays real-world websites in a testbed subject to different protocols and network settings.
\end{itemize}
\vspace{-0.55em}
\afblock{Structure.}
\sref{sec:background} examines related work, highlights the evaluation metrics and introduces to the Mahimahi framework.
\sref{sec:testbed} explains our testbed, network configuration, and protocol considerations.
\sref{sec:evaluation} shows the results of the measurement.
Finally, \sref{sec:conclusion} concludes this paper.

\vspace{-0.5em}
\section{Related Work and Background}
\label{sec:background}
QUIC is subject to a body of studies~\cite{biswal16:globecom:does, carlucci15:sac:http, cook17:icc:quic, kakhki17:imc:taking, megyesi16:icc:quick, yu17:ipccc:quic, nepomuceno18:iscc:quic}, most compare QUIC against some combination of TCP+TLS+\-HTTP/1.1 or HTTP/2.
But to the best of our knowledge, all use stock TCP configurations measuring a likely unoptimized TCP version to a QUIC version that inherently contains available TCP optimizations.
Yu et al.~\cite{yu17:ipccc:quic} is the only study on the impact of packet pacing for QUIC as a tuning option.
However, no further comparison to TCP is made.

Generally, the related work can be divided into two categories depending on their measurement approach.
One body of research~\cite{seufert:qomex19:quic, cook17:icc:quic, megyesi16:icc:quick} measures against websites hosted on public servers utilizing both QUIC and TCP---however, usually operated by Google.
Thus, they do not have any access to the servers, which makes tuning the protocol impossible and the configurations in use are unknown.
The second body~\cite{biswal16:globecom:does, carlucci15:sac:http, kakhki17:imc:taking, nepomuceno18:iscc:quic} uses self-hosted servers, in principle allowing for tuning, however, none of them does so.

One critical difference between TCP and QUIC is their connection establishment since QUIC by design needs fewer RTTs than traditional TCP+TLS until actual website payload can be exchanged.
Cook et al.~\cite{cook17:icc:quic} already take into account that there is a difference between first and repeated connections that require each one less RTT for both protocols.
Nevertheless, QUIC still has a one RTT advantage in both connections, repeated as well as first, and again this fact is not dealt with any further.

Since today's websites consist of various resources hosted by several providers, many connections to different servers are established even for fetching a single website.
Many studies consider websites with varying resources but deployed by a single server only~\cite{biswal16:globecom:does, carlucci15:sac:http, megyesi16:icc:quick}.
To study realistic Web sites, the Mahimahi framework~\cite{netravali15:atc:mahimahi} was designed to replicate this multi-server nature of current websites into a testbed (see \sref{sec:mahimahi}).
Nepomuceno et al.~\cite{nepomuceno18:iscc:quic} perform a study with Mahimahi but find that QUIC is outperformed by TCP which does not coincide with our and related work.
We believe this is due to the use of the Caddy QUIC server, which is known to not (yet) perform very well~\cite{Caddy}.
Also, they did not configure any bandwidth limitations.

\subsection{Web Performance Metrics}
\label{sec:metrics}

We aim to evaluate the performance of a different protocol stack on a broad set of standard Web performance metrics.
Besides network characteristics like goodput or link utilization as measured in~\cite{carlucci15:sac:http, yu17:ipccc:quic}, \textit{Page Load Time} (PLT) is the most used metric.
But PLT does not always match user-perceived performance~\cite{kelton17:nsdi:webgaze,zimmermann17:inetqoe:qoepush,bocchi17:pam:web}, \eg{} it includes the loading performance of \textit{below}-the-fold content that is not displayed and thus not reflected in end-user perception.
This is why we decide to focus more closely on state-of-the-art visual metrics that are known to better correlate with human perception.
These metrics are derived from video recordings of the pages loading process \textit{above}-the-fold as recommended by~\cite{brutlag11:wpoc:above,gao17:inetqoe:perceived}.

Metrics of interest are the time of the \textit{First Visual Change} (FVC), \textit{Last Visual Change} (LVC), and time the website reaches visual completeness of a desired threshold in percent.
In our case, \textit{Visual Complete 85} (VC85), which corresponds to the point in time measured from the navigation start when the currently rendered website's above-the-fold matches to $85\%$ the final website picture.
Only navigation start can be used as start point since visual metrics are derived from video recordings only (see \sref{sec:testbed:mahimahi} how we deal with DNS impacting the measurement).
Lastly, we also take into account the \textit{Speed Index} (SI)~\cite{SpeedIndex}.

\subsection{Website Replay with Mahimahi}
\label{sec:mahimahi}

Mahimahi~\cite{netravali15:atc:mahimahi} is a framework designed to replicate real-world websites with their multi-server structure in a testbed.
It uses HTTP traffic recordings that are later replayed.
Mahi\-mahi preserves the multi-server nature with the help of virtualized Web servers.
Mahimahi is built upon multiple shell commands that can be stacked to create a virtual network.
Each shell allows for modifying a single aspect of the traversing network flow, \eg{} generating loss or limiting the bandwidth.
Mahimahi yields realistic conditions for performance measurements~\cite{netravali15:atc:mahimahi}.
This way, it enables repeatable and controllable studies with real-world websites.

\section{Testbed Setup}
\label{sec:testbed}
We will now continue to explain how we design our testbed to perform eye-level comparisons of TCP and QUIC.

\subsection{Selecting and Recording Websites}
\label{sec:testbed:doamins}

\begin{figure}
	\centering
	\includegraphics{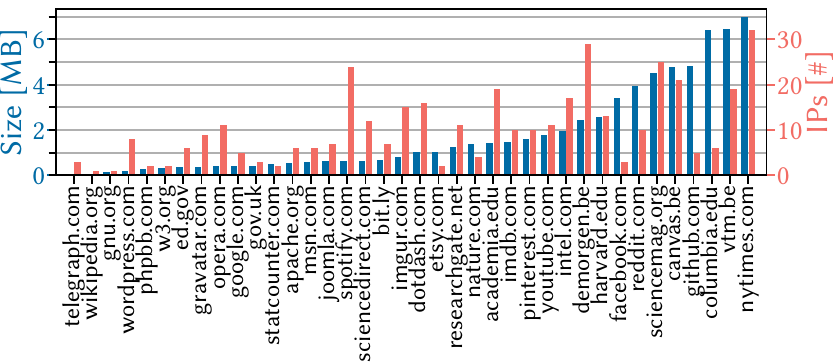}
	\vspace{-2em}
	\caption{This figure depicts the download size of the replayed websites (blue) and the number of unique IP addresses that need to be contacted for resources (red).}
	\label{fig:testbed:url-sizes}
	\vspace{-1em}
\end{figure}

\afblock{Websites.}
We want to choose websites that replicate a real-world picture of commonly used websites.
The goal is to obtain a small set of domains diverse in size, resources, and involved servers.
As there is no standard test set of such website, we use the domain collection from~\cite{wijnants2018http} consisting of $40$ different websites from which we had to exclude two.
One domain is a private project website and the other failed to record and reply properly.
The domains originate from the Alexa~\cite{Alexa} and Moz~\cite{Moz} ranking lists and were chosen in a way to obtain a good distribution of page size and resource counts~\cite{wijnants2018http}, see \fig{fig:testbed:url-sizes}.
The bars in red suggest the majority of our tested sites to use multi-server infrastructures, highlighting the relevance of replicating it with Mahimahi.

\afblock{Recording.}
Downloading of the websites was not performed with the tool provided by Mahimahi.
Instead, we utilize Mitmproxy with a custom script that dumps the raw HTTP responses of the server to disk.
According to the Google QUIC server specification~\cite{QuicServer} the \textit{transfer-encoding} header must be removed if its value is \textit{chuncked}.
The same holds for the \textit{alternate-protocol} header for any value.
Other than that, the recorded HTTP responses remain unchanged.

In post-processing, few resource files needed to be downloaded additionally, since, \eg{} the header of the github.com website loads a random image from a fixed collection via JavaScript.

\subsection{Replaying with Mahimahi}
\label{sec:testbed:mahimahi}

\afblock{Mahimahi.}
To support a state-of-the-art QUIC in Mahimahi, we include Google's QUIC server from the Chromium sources utilizing QUIC Version 43.
For TLS1.3 and HTTP/2, we replace the Mahimahi default Apache server with NGINX.
All NGINX servers forward the requests to a single uWSGI proxy server that provides the previously recorded HTTP responses from main memory.
Similarly, the QUIC servers use their built-in feature loading all responses from a folder into memory.
Finally, we create a self-signed certificate authority (CA) and incorporate it to the Chrome browser's list of trusted CAs to circumvent any authentication errors.

\afblock{Enforcing QUIC or TCP+TLS1.3+H2.}
We want to be sure that only QUIC or TCP is used.
On the one hand, we accomplish this using Chrome flags, to enforce QUIC, we set ``--enable-quic --origin-to-force-quic-on=*'' and ``--disable-quic''  for TCP respectively.
On the other hand, the QUIC and NGINX servers never run at the same time.
In the TCP case, each request is performed over TLS1.3 and HTTP/2.
There are no resources that get transmitted unencrypted.

\afblock{Protocol Tuning.}
To allow for a fair comparison between TCP and QUIC, we tune the stock TCP stack of a Linux kernel to more closely match QUIC's defaults.
This is done by increasing the initial window to 32 segments, enabling pacing, setting no slow start after idle and tuning the kernel buffers.
QUIC by default also uses an initial window of 32 and pacing.
Since we expect the employed congestion control algorithm to significantly impact the measured performance, we incorporated one scenario for TCP and QUIC each utilizing BBR~\cite{Cardwell:BBR} instead of CUBIC~\cite{Ha:CUBIC}.
An overview of the five protocol configurations is shown in \tab{tab:testbed:protocols}.

\begin{table}[t]
	\centering
	\begin{tabular}{@{}l|l@{}}
		\hline
		\ \textbf{Protocol} & \textbf{Description} \\ \hline\hline
		\ TCP & Stock TCP (Linux): IW10, Cubic \\ \hline
		\ TCP+ & \begin{tabular}[c]{@{}l@{}}\unit[IW]{32}, Pacing, Cubic, tuned buffers,\\ no slow start after idle\end{tabular} \\ \hline
		\ TCP+BBR & TCP+, but with BBR as congestion control \\ \hline
		\ QUIC & Stock Google QUIC: \unit[IW]{32}, Pacing, Cubic \\ \hline
		\ QUIC+BBR & QUIC, but with BBR as congestion control \\ \hline
	\end{tabular}
	\caption{Protocol configuration used in our tests.}
	\label{tab:testbed:protocols}
	\vspace{-2em}
\end{table}

TCP Fast Open~\cite{radhakrishnan11:conext:TFO} and TLS1.3 early-data~\cite{RFC8446} are two possible options to tune TCP/TLS even further.
We decided against both techniques because of the following reasons.
TLS1.3 early-data was not supported by the Chrome browser at the time of the measurement and as it is prone to replay attacks requires idempotency which further challenges its widespread use.
TCP Fast Open is not widely deployed on the Internet today~\cite{mandalari2015tcp, paasch2016network}.
Moreover, we always measure the website performance with a fresh browser and clean caches, thus QUIC has to perform an extra RTT for connection establishment as well and does not use $0$-RTT connections.

\afblock{Network Settings.}
For network emulation, the built-in tools from Mahimahi are used.
We stack the following three network parameters from server to client with Mahimahi shells.
First, a packet gets delayed in either direction, both adding up to the desired minimum latency.
Second, the link shell implements a drop-tail buffer limiting the throughput per direction.
Finally, the loss shell drops packets at random for both directions equally.
The loss is configured, such that the chance for two packets, \eg{} request and response, getting transmitted successfully equals $1-p$ with $p$ being the desired loss rate.
The implemented values are shown in \tab{tab:testbed:network}.
\begin{table}
	\centering
	\begin{tabular}{l|r|r|r|r} 
		\textbf{Network} & \textbf{Uplink} & \textbf{Downlink} & \textbf{Delay} & \textbf{Loss} \\ \hline \hline
		DSL		&    \unit[5]{Mbps} &   \unit[25]{Mbps} &  \unit[24]{ms} &  \unit[0.0]{\%} \\ \hline
		LTE		&  \unit[2.8]{Mbps} & \unit[10.5]{Mbps} &  \unit[74]{ms} &  \unit[0.0]{\%} \\ \hline
		DA2GC	& \unit[.468]{Mbps} & \unit[.468]{Mbps} & \unit[262]{ms} &  \unit[3.3]{\%} \\ \hline
		MSS		& \unit[1.89]{Mbps} & \unit[1.89]{Mbps} & \unit[760]{ms}&  \unit[6.0]{\%} \\ 
	\end{tabular}
	\caption{Network configurations. Queue size is set to \unit[200]{ms} except for DSL with \unit[12]{ms}.}
	\label{tab:testbed:network}
	\vspace{-1em}
\end{table}
Bandwidth and delay values for DSL and LTE are taken from~\cite{Breitbandmessung}, we assume no additional loss here.
The last two networks emulate slow links measured from in-flight WLAN services~\cite{rula2018mile}.
Except for the DSL link with $12$ms maximal queueing delay, we assume rather bloated buffers of \unit[200]{ms}.
Thus, our configured delay is the minimum delay and queuing further adds jitter up to the configured buffer size.

\afblock{Validation.}
Before conducting measurements, we validate the implemented testbed regarding the network and protocol parameters ensuring the correct protocol choice.
We found that the Chromium browser's DNS timeout of \unit[5]{s} significantly distorts a measurement when a DNS packet is lost and thus moved the DNS server such that no traffic shaping is applied to DNS traffic.
\begin{figure}
	\centering
	\includegraphics{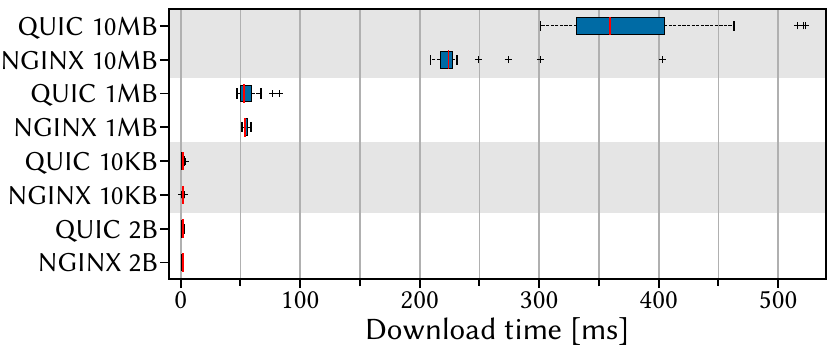}
	\vspace{-2em}
	\caption{Boxplot of server download speeds in our testbed (31 repetitions and no bandwidth limitation).}
	\label{fig:testbed:server-speed}
\end{figure}
Moreover, \fig{fig:testbed:server-speed} shows that both server variants yield similar performance for files $\le$ \unit[1]{MB}.
This suggests that our results are not biased by the servers' implementations.
For this test, we repeated 31 downloads of a single file with the Chromium browser under static network conditions---only \unit[10]{ms} minimum delay, no loss, and no bandwidth limits.
The gap between NGINX and QUIC server emerging at a file size of \unit[10]{MB} is not relevant since our website sizes are much smaller (see \fig{fig:testbed:url-sizes}).
Independent resources are even smaller, the largest being \unit[4]{MB}.

\begin{figure*}
	\centering
	\includegraphics{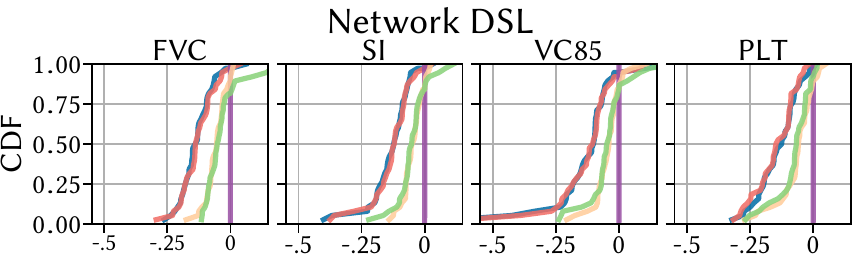}\hfill\includegraphics{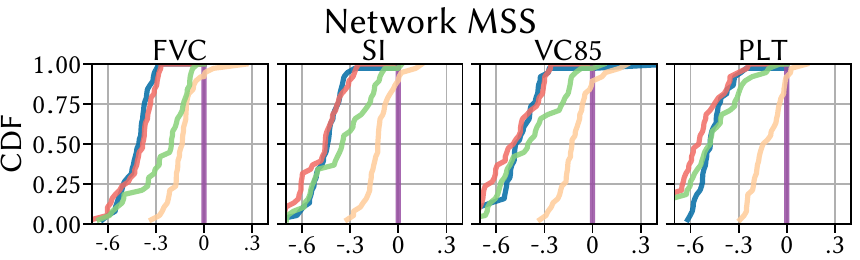}\\
	\includegraphics{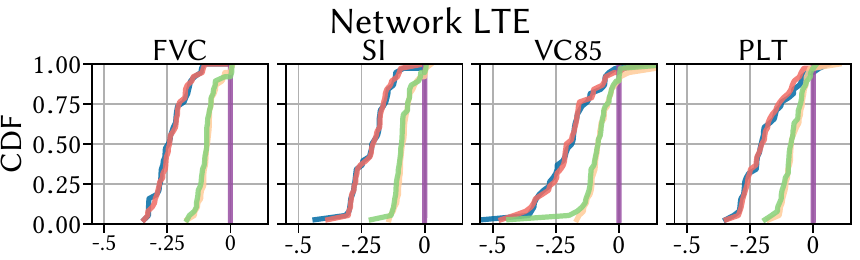}\hfill\includegraphics{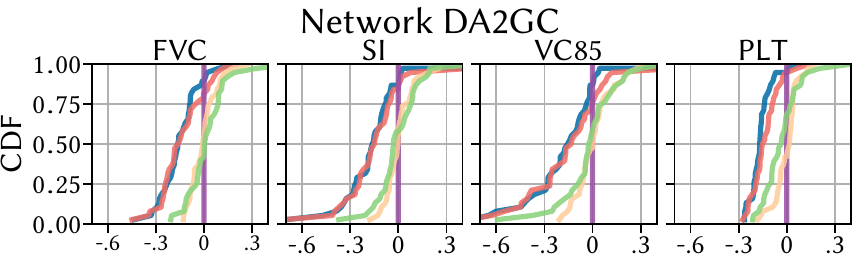}\\
	\includegraphics{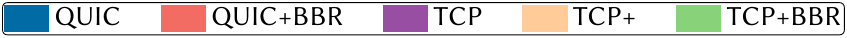}
	\vspace{-0.5em}
	\caption{CDF of the performance gain over all websites with TCP as reference protocol. If the performance gain is $<0$ (left side of plot) then the compared protocol is faster than TCP.}
	\label{fig:eval:faster}
\end{figure*}

\subsection{Performing Measurements}
The actual measurements are performed inside a virtual machine equipped with 6 cores and \unit[8]{GB} of memory running Arch Linux kernel Version 4.18.16.
To measure a single setting consisting of one website, network, and protocol configuration, a Mahimahi replay shell with the described network stack is used.
A single setting gets measured over $31$ runs to gain statistical significance and at the same time keep the number of runs/videos manageable.
We utilize the Browsertime~\cite{browsertime} framework to instrument the browser.
It records videos of the loading process that we subsequently evaluate for the visual metrics.
For each run, Browsertime opens up a fresh Chromium browser Version 70.0.3538.77.
In total, this leads to 760 configurations (38 domains, 4 network, and 5 protocol settings).
We validated that each run completed successfully by reviewing the video recordings manually.

\section{QUIC vs. TCP Performance}
\label{sec:evaluation}

We evaluate the performance difference with all metrics in the different network settings (across all tested websites) by means of a performance gain.
The following equation explains the calculation of the performance gain between a reference protocol, \eg{} TCP and a protocol to compare with like QUIC.
$\overline{X}$ correspond to the mean over the $31$ runs.
\begin{equation*}
	\textit{performance gain}_{QUIC}^{TCP} = \frac{\overline{X}_{QUIC} - \overline{X}_{TCP}}{\overline{X}_{TCP}}
\end{equation*}
If not stated otherwise, numbers provided in the text are mean performance gains over all websites for SI.
Besides comparing means we also utilize an ANOVA test to tell whether there is a statistically significant difference in the distribution of the $31$ runs of two protocols.
If the ANOVA test for two settings is $p<0.01$ (significance level), we count the setting with the lower mean as significantly faster otherwise no conclusion can be drawn.
The results of our measurements are depicted in \fig{fig:eval:faster}.
We show the CDFs of the performance gain of the different metrics comparing stock TCP to the other protocol stacks.
LVC is left out in this figure because in contrast to PLT there is no relevant difference visible.

\newcommand{\hardnumber}[3]{#1\textsupsub{\tiny #3}{\tiny #2}}

\afblock{DSL and LTE.}
For the lossless DSL and LTE scenarios, the protocols separate into two groups both yielding similar performance gains.
TCP+ (DSL: \hardnumber{\MeanGainSiDslTcpTcpp}{TCP}{TCP+}, LTE: \hardnumber{\MeanGainSiLteTcpTcpp}{TCP}{TCP+}) and TCP+BBR (DSL: \hardnumber{\MeanGainSiDslTcpTcpbbr}{TCP}{TCP+BBR}, LTE: \hardnumber{\MeanGainSiLteTcpTcpbbr}{TCP}{TCP+BBR}) perform almost indistinguishable but against TCP, there is a noticeable improvement visible throughout all metrics.
Similarly, QUIC (DSL: \hardnumber{\MeanGainSiDslTcppQuic}{TCP+}{QUIC}, LTE: \hardnumber{\MeanGainSiLteTcppQuic}{TCP+}{QUIC}) and QUIC+BBR (DSL: \hardnumber{\MeanGainSiDslTcpbbrQuicbbr}{TCP+BBR}{QUIC+BBR}, LTE: \hardnumber{\MeanGainSiLteTcpbbrQuicbbr}{TCP+BBR}{QUIC+BBR}) perform equally but are still quite a bit faster than the two tuned TCP variants.
For these two networks, the congestion control choice does not make a significant difference, which is likely due to the small queue.
Stock TCP indeed lags behind all other protocols showing that stock TCP should not be used to compare against QUIC here.
QUIC achieves to decrease the average SI by \hardnumber{\unit[\MeanDiffSiDslTcpQuic]{ms}}{TCP}{QUIC} (DSL) and \hardnumber{\unit[\MeanDiffSiLteTcpQuic]{ms}}{TCP}{QUIC} (LTE), but also against TCP+ by still \hardnumber{\unit[\MeanDiffSiDslTcppQuic]{ms}}{TCP+}{QUIC} (DSL) and \hardnumber{\unit{\MeanDiffSiLteTcppQuic}{ms}}{TCP+}{QUIC} (LTE).

In a second step, we take a look at the ANOVA test results focussing on DSL (LTE yields equivalent results).
When comparing the runs of TCP+ and QUIC in DSL with PLT as the metric with each other, $30$ of the $38$ websites yield a significant improvement with QUIC.
For the remaining $8$ websites, none was significantly faster than TCP+.
For SI even $32$ are faster and only $6$ show no significant difference.
Similar results can be seen when comparing QUIC+BBR with TCP+BBR this way.
For TCP+ and TCP in the same scenario with PLT as the metric, $25$ websites are faster with TCP+, for $12$ there is no significant difference and only $1$ website was significantly slower.
Again when comparing TCP+BBR with TCP+ and similarly QUIC+BBR with QUIC for DSL and LTE throughout all metrics, we find for the majority of the websites no difference.
These results line up with the results shown in \fig{fig:eval:faster}.
Moreover, the steep incline of the CDFs for QUIC and TCP+ indicate that the website size or structure seems to have little influence on the achievable gain.
Only looking at SI and VC85, we see a small percentage of measurements where QUIC has a significantly higher gain.

\afblock{In-Flight Wifi.}
For the networks MSS and DA2GC, the overall picture is quite similar---meaning QUIC as well as QUIC+BBR, are usually faster than TCP+ (MSS: \hardnumber{\MeanGainSiMssTcppQuic}{TCP+}{QUIC}, DA2GC: \hardnumber{\MeanGainSiDaagcTcppQuic}{TCP+}{QUIC}) and TCP+BBR (MSS: \hardnumber{\MeanGainSiMssTcpbbrQuicbbr}{TCP+BBR}{QUIC+BRR}, DA2GC:  \hardnumber{\MeanGainSiDaagcTcpbbrQuicbbr}{TCP+BBR}{QUIC+BBR}).
But there are some important differences, for the MSS link with the highest loss rate (6\,\%), TCP+BBR operates much better than TCP+ (\hardnumber{\MeanGainSiMssTcppTcpbbr}{TCP+}{TCP+BBR}). 
Since BBR does not use loss as a congestion signal it increases its rate regardless of this random loss.
This means that in this case, the choice in congestion control has a greater impact on the performance than the protocol choice itself.
At the time of the FVC, TCP+BBR is already \unit[\MeanDiffFvcMssTcppTcpbbr]{ms} (avg.) quicker than TCP+ but with each later metric, the gap widens so that at PLT, TCP+BBR can keep up the pace even against QUIC and is \unit[\MeanDiffPltMssTcpbbrQuic]{ms} (\MeanGainPltMssTcpbbrQuic$\times$) quicker.
This shows that TCP with BBR needs some time to catch up and thus affects the FVC much more than the later PLT.
For the QUIC protocols, the picture is similar.
At first, QUIC and QUIC+BBR are similarly fast and mostly better than TCP+BBR.
But as the loading process commences QUIC+BBR outperforms QUIC slightly, \eg{} \hardnumber{\unit[\MeanDiffSiMssQuicQuicbbr]{ms}}{QUIC}{QUIC+BBR} better SI.
QUIC with CUBIC, nevertheless, is reasonably fast being still a legit option to use.
The shape of the performance gain CDFs of QUIC+BBR and TCP+BBR are very similar especially for PLT highlighting the influence of the congestion control once again.
We believe that QUIC with CUBIC is still competitive due to QUIC's ability to circumvent head-of-line blocking and its large SACK ranges.
For the MSS network, QUIC reduces the SI by \hardnumber{\unit[\MeanDiffSiMssTcppQuic]{ms}}{TCP+}{QUIC} (avg.) compared to TCP+ and by \hardnumber{\unit[\MeanDiffSiMssTcpbbrQuicbbr]{ms}}{TCP+BBR}{QUIC+BBR}when taking both BBR protocols into account.

The last network, DA2GC, also has a high loss rate (3.3\,\%) but a much lower bandwidth.
This is the only scenario where we observe no significant difference for most websites among all TCP configurations even with the ANOVA test.
We also see that in a small fraction of our measurements stock TCP outperforms QUIC and the tuned TCP variants.
Nevertheless, again the QUIC variants are generally significantly faster with a higher performance gain at the FVC (\eg{} \hardnumber{\MeanGainFvcDaagcTcppQuic}{TCP+}{QUIC}) that persists towards the PLT (\eg{} \hardnumber{\MeanGainPltDaagcTcppQuic}{TCP+}{QUIC}).
The choice of the congestion control algorithm does not seem to have a significant impact here likely due to the low bandwidth.
Only for PLT we find QUIC with CUBIC to be slightly superior over QUIC with BBR.
There is not a single website where QUIC+BBR yields a significantly faster performance.
The SI decreases with QUIC by \hardnumber{\unit[\MeanDiffSiDaagcTcppQuic]{ms}}{TCP+}{QUIC} vs.\ TCP+ and by \hardnumber{\unit[\MeanDiffSiDaagcTcpbbrQuicbbr]{ms}}{TCP+BRR}{QUIC+BBR} for BBR.

\afblock{Discussing Metrics.}
Some of the websites exhibit very poor performance regarding the visual metrics VC85 and SI.
We observe this behavior especially for the DA2GC network with performance gains of up to $+1.0$ compared to stock TCP (not shown, plots cropped for readability).
The reason for these outliers is that the protocol choice has such a substantial impact on some websites that their resources load in different orders resulting in very distinct rendering sequences.

\begin{figure}
	\centering
	\includegraphics[width=\columnwidth]{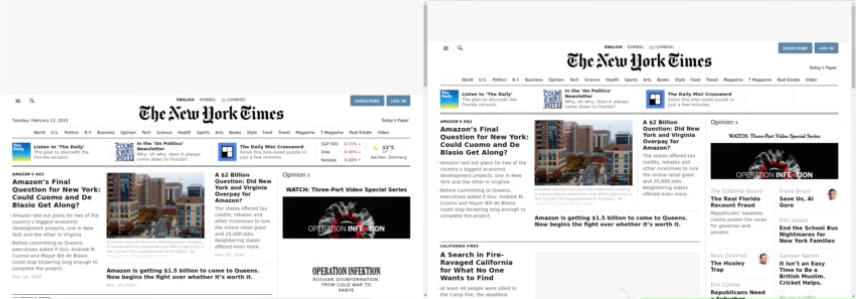}
	\vspace{-1.5em}
	\caption{Screenshot during the loading process of the nytimes.com website. Left TCP right QUIC. QUIC in comparison delays a top banner leading to bad scores in visual metrics compared to the final website.}
	\label{fig:nytimes}
\end{figure}

\fig{fig:nytimes} shows such a scenario exemplary for the nytimes.com website in the DA2GC network.
Here TCP reaches VC85 after ${\sim}48$s whereas QUIC needs ${\sim}124$s even though the PLT for QUIC (${\sim}141$s) is much faster than for TCP (${\sim}170$s).
For TCP the upper part of the website loads comparably early such that the lower elements are already rendered at their final positions.
In contrast to that QUIC manages to receive the lower contents first.
Later, when the upper banner completes loading it shifts the whole website down.
Therefore, VC85 fails to express this setting given the large shift.
Similarly, SI is affected since it integrates over visual completeness over time.
Thus, it critically depends on the website, the browser's loading order, and a user's preference for how a website should load to know which metric to use.

\afblock{Protocol Design Impact.}
Within our testbed, any TCP configuration needs to fulfill two complete RTTs before the actual HTTP request can be sent out to the server---TCP handshake plus TLS setup.
In contrast, QUIC requires only one RTT to do so---the first CHLO gets rejected by the server since the server certificates are unknown to the client.
We are interested in whether this 1 RTT difference can explain the remaining performance gap between QUIC and TCP+.
However, the complex interactions with multiple servers complicate an analysis since these connections are interleaved simply subtracting 1 RTT is not possible.
We, therefore, take a look at two websites served only via a single IP (see \fig{fig:testbed:url-sizes}): wikipedia.org and gnu.org.
We subtract one RTT from the FVC, as the earliest metric and one RTT from the PLT as the latest completing metric.
\tab{tab:rtt} shows the results in the different network settings for TCP+ and QUIC and additionally for MSS using the BBR variants of both.

\begin{table}
	\Small
	\renewcommand{\arraystretch}{0.8}
	\begin{tabular}{ccccc}
	\multicolumn{1}{c|}{\textbf{Net}} & \multicolumn{1}{c|}{\textbf{Website}} & \multicolumn{1}{c|}{\textbf{Metric}} & \multicolumn{1}{c|}{\textbf{[ms]}} & \multicolumn{1}{c}{\textbf{[RTT]}} \\ \hline \hline
	\multicolumn{1}{l|}{DSL} & \multicolumn{1}{l|}{gnu.org} & \multicolumn{1}{l|}{FVC} & \multicolumn{1}{r|}{0.5} & \multicolumn{1}{r}{0.020} \\
	\multicolumn{1}{l|}{DSL} & \multicolumn{1}{l|}{wikipedia.org} & \multicolumn{1}{l|}{FVC} & \multicolumn{1}{r|}{-8.2} & \multicolumn{1}{r}{-0.341} \\
	\multicolumn{1}{l|}{DSL} & \multicolumn{1}{l|}{gnu.org} & \multicolumn{1}{l|}{PLT} & \multicolumn{1}{r|}{1.6} & \multicolumn{1}{r}{0.066} \\
	\multicolumn{1}{l|}{DSL} & \multicolumn{1}{l|}{wikipedia.org} & \multicolumn{1}{l|}{PLT} & \multicolumn{1}{r|}{-3.1} & \multicolumn{1}{r}{-0.128} \\ \hline
	\multicolumn{1}{l|}{LTE} & \multicolumn{1}{l|}{gnu.org} & \multicolumn{1}{l|}{FVC} & \multicolumn{1}{r|}{0.6} & \multicolumn{1}{r}{0.008} \\
	\multicolumn{1}{l|}{LTE} & \multicolumn{1}{l|}{wikipedia.org} & \multicolumn{1}{l|}{FVC} & \multicolumn{1}{r|}{-40} & \multicolumn{1}{r}{-0.538} \\
	\multicolumn{1}{l|}{LTE} & \multicolumn{1}{l|}{gnu.org} & \multicolumn{1}{l|}{PLT} & \multicolumn{1}{r|}{-30} & \multicolumn{1}{r}{-0.412} \\
	\multicolumn{1}{l|}{LTE} & \multicolumn{1}{l|}{wikipedia.org} & \multicolumn{1}{l|}{PLT} & \multicolumn{1}{r|}{-13} & \multicolumn{1}{r}{-0.175} \\ \hline
	\multicolumn{1}{l|}{MSS} & \multicolumn{1}{l|}{gnu.org} &  \multicolumn{1}{l|}{FVC} & \multicolumn{1}{r|}{-196} & \multicolumn{1}{r}{-0.258} \\
	\multicolumn{1}{l|}{MSS} & \multicolumn{1}{l|}{wikipedia.org} &  \multicolumn{1}{l|}{FVC} & \multicolumn{1}{r|}{-412} & \multicolumn{1}{r}{-0.542} \\
	\multicolumn{1}{l|}{MSS} & \multicolumn{1}{l|}{gnu.org} & \multicolumn{1}{l|}{PLT} & \multicolumn{1}{r|}{-1100} & \multicolumn{1}{r}{-1.447} \\
	\multicolumn{1}{l|}{MSS} & \multicolumn{1}{l|}{wikipedia.org} & \multicolumn{1}{l|}{PLT} & \multicolumn{1}{r|}{-529} & \multicolumn{1}{r}{-0.696} \\ \hline
	\multicolumn{1}{l|}{DA2GC} & \multicolumn{1}{l|}{gnu.org} &  \multicolumn{1}{l|}{FVC} & \multicolumn{1}{r|}{-130} & \multicolumn{1}{r}{-0.497} \\
	\multicolumn{1}{l|}{DA2GC} & \multicolumn{1}{l|}{wikipedia.org} &  \multicolumn{1}{l|}{FVC} & \multicolumn{1}{r|}{-1384} & \multicolumn{1}{r}{-5.283} \\
	\multicolumn{1}{l|}{DA2GC} & \multicolumn{1}{l|}{gnu.org} & \multicolumn{1}{l|}{PLT} & \multicolumn{1}{r|}{39} & \multicolumn{1}{r}{0.150} \\
	\multicolumn{1}{l|}{DA2GC} & \multicolumn{1}{l|}{wikipedia.org} & \multicolumn{1}{l|}{PLT} & \multicolumn{1}{r|}{-1005} & \multicolumn{1}{r}{-3.834} \\ \hline \hline
	\multicolumn{1}{l|}{MSS} & \multicolumn{1}{l|}{gnu.org} & \multicolumn{1}{l|}{FVC} & \multicolumn{1}{r|}{-404} & \multicolumn{1}{r}{-0.532} \\
	\multicolumn{1}{l|}{MSS} & \multicolumn{1}{l|}{wikipedia.org} & \multicolumn{1}{l|}{FVC} & \multicolumn{1}{r|}{-143} & \multicolumn{1}{r}{-0.189} \\
	\multicolumn{1}{l|}{MSS} & \multicolumn{1}{l|}{gnu.org} & \multicolumn{1}{l|}{PLT} & \multicolumn{1}{r|}{-477} & \multicolumn{1}{r}{-0.628} \\
	\multicolumn{1}{l|}{MSS} & \multicolumn{1}{l|}{wikipedia.org} & \multicolumn{1}{l|}{PLT} & \multicolumn{1}{r|}{451} & \multicolumn{1}{r}{0.593} \\ \hline
	\end{tabular}

	\caption{Difference between the means over the 31 runs of QUIC and TCP+ when subtracting one RTT. Values $<$0 denote that QUIC was faster. The lower MSS table compares QUIC+BBR and TCP+BBR.}
	\label{tab:rtt}
	\vspace{-1.9em}
\end{table}

For DSL and LTE the corrected difference is below one RTT and there are three cases where even TCP+ is slightly faster now.
For MSS in all cases with CUBIC as the congestion control, QUIC is faster but only to a maximum of $1.4\times$ RTT.
Since within this network congestion control has a huge impact, we consider also BBR here.
Overall in MSS with BBR, the difference is also below of one RTT and for wikipedia.org and PLT even TCP+BBR is faster.
Instead with DA2GC, the outcome is clearly for QUIC for the wikipedia.org website.
\tab{tab:rtt} shows nicely that QUIC's RTT reducing design clearly improves the performance.
Even though, TCP Fast Open and TLS 1.3 early-data would close the gap, especially Fast Open remains challenging to deploy.
Furthermore, having no head-of-line blocking could still be a reason why in the majority of the cases QUIC is still slightly faster, especially, when the networks are lossy.
We expect further improvements when using 0-RTT connection establishment with QUIC.

\section{Conclusion}
\label{sec:conclusion}

Comparisons between TCP and QUIC have often been biased up until now.
In this paper, we extended the Mahimahi framework to support QUIC and perform reproducible performance measurements of $38$ websites under different protocol and network scenarios.
We show that tuning TCP parameters has a tremendous impact on the results for performance comparisons which can not be neglected when comparing TCP and QUIC.
Yet, in many settings, QUIC's performance is still superior but the gap gets narrower.
Moreover, we find that QUIC's higher performance is caused mostly due to its superior design during the connection establishment.
We assume that besides the RTT reducing design, features like no head-of-line blocking increase QUIC's performance, especially in lossy networks. 
In those lossy networks, we also find that the choice of the congestion control algorithm has a much larger impact than the protocol itself.
In our opinion, QUIC is still the preferred protocol for the future Web since it paves the way for continuous evolution.

 \clearpage

\begin{acks}
This work has been funded by the DFG as part of the CRC 1053 MAKI and SPP 1914 REFLEXES.
\end{acks}

\bibliographystyle{ACM-Reference-Format}
\balance
\bibliography{paper}
\end{document}